\documentclass[prb,showpacs,citeautoscript]{revtex4}
\usepackage{graphicx}
\usepackage{amsmath}
\graphicspath{{./img/}}
\newcommand{\T}{\mathcal{T}}
\begin{document}
\title{Resonant laser excitation of molecular wires}
\author{Sigmund Kohler,
        J\"org Lehmann,
        S\'ebastien Camalet,
        Peter H\"anggi}
\affiliation{Institut f\"ur Physik, Universit\"at Augsburg,
        Universit\"atsstra\ss e~1, D-86135 Augsburg, Germany}
\date{September 17, 2002}
%
\begin{abstract}
We investigate the influence of external laser excitations on the average
current through bridged molecules.  For the computation of the current, we use
a numerically very efficient formalism that is based on the Floquet solutions
of the time-dependent molecule Hamiltonian.  It is found that the current as a
function of the laser frequency exhibits characteristic peaks originating from
resonant excitations of electrons to bridge levels which are unoccupied in the
absence of the radiation.  The electrical current through the molecule can
exhibit a drastic enhancement by several orders of magnitude.
\pacs{
85.65.+h, 
33.80.-b, 
73.63.-b, 
05.60.Gg 
}
\end{abstract}
\maketitle

\section{Introduction}

In a seminal work \cite{Aviram1974a}, Aviram and Ratner proposed almost thirty
years ago to build elements of electronic circuits --- in their case a
rectifier --- with single molecules.  In the present days their vision starts
to become reality and the experimental and theoretical study of such systems
enjoys a vivid activity \cite{Joachim2000a,Nitzan2001a,Hanggi2002a}.
Recent experimental progress has enabled reproducible measurements
\cite{Cui2001a,Reichert2002a} of weak tunneling currents through molecules
which are coupled by chemisorbed thiol groups to the gold surface of external
leads.  A necessary ingredient for future technological applications will be
the possibility to control the tunneling current through the molecule.

Typical energy scales in molecules are in the optical and the infrared regime,
where basically all of the today's lasers operate. Hence, lasers represent a
natural possibility to control atoms or molecules and also currents
through them.  It is for example possible to induce by the laser field an
oscillating current in the molecule which under certain asymmetry conditions is
rectified by the molecule resulting in a directed electron transport even in
the absence of any applied voltage \cite{Lehmann2002b,Lehmann2002d}.  Another
theoretically predicted effect is the current suppression by the laser field
\cite{Lehmann2002c} which offers the possibility to control and switch the
electron transport by light. 
Since the considered frequencies lie below typical plasma frequencies
of metals, the laser light will be reflected at the metal surface,
\textit{i.e.}  it does not penetrate the leads. Consequently, we do
not expect major changes of the leads' bulk properties --- in
particular each lead remains close to equilibrium.  Thus, to a good
approximation, it is sufficient to consider the influence of the
driving solely in the molecule Hamiltonian.  In addition, the energy
of infrared light quanta is by far smaller than the work function of a
common metal, which is of the order of $5\,\mathrm{eV}$.  This
prevents the generation of a photo current, which otherwise would
dominate the effects discussed below.

Recent theoretical descriptions of the molecular conductivity in
non-driven situations are based on a scattering approach
\cite{Mujica1994a,Datta1995a}, or assume that the underlying transport
mechanism is an electron transfer reaction from the donor to the
acceptor site and that the conductivity can be derived from the
corresponding reaction rates \cite{Nitzan2001a}.  It has been
demonstrated that both approaches yield essentially identical results
in a large parameter regime \cite{Nitzan2001b}.  Within a
high-temperature limit, the electron transport on the wire can be
described by inelastic hopping events \cite{Petrov2001a,Lehmann2002a}.

Atoms and molecules in strong oscillating fields have been widely studied
within a Floquet formalism \cite{Manakov1986a,Grifoni1998a}.  This suggests
utilizing the tools that have been acquired in that area, thus,  developing a
transport formalism that combines Floquet theory for a driven molecule with
the many-particle description of transport through a system that is coupled to
ideal leads \cite{Lehmann2002b, Lehmann2002c, Lehmann2002d}.  Such an approach is
devised much in the spirit of the Floquet-Markov theory
\cite{Blumel1989a,Kohler1997a} for driven dissipative quantum systems.


\section{Floquet treatment of the electron transport}
\label{sec:formalism}
\begin{figure}
\includegraphics[width=.6\columnwidth]{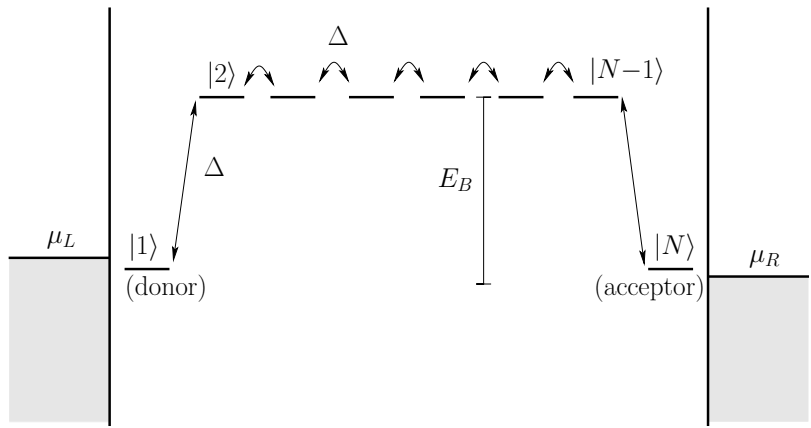}
\caption{\label{fig:wire}
Level structure of a molecular wire with $N=8$ atomic sites
which are attached to two leads.
}
\end{figure}

The entire system of the wire in the laser-field, the leads, and the molecule-lead
coupling as sketched in Figure~\ref{fig:wire} is described by the Hamiltonian
\begin{equation}
\label{wire-lead-hamiltonian}
H(t)=H_{\text{wire}}(t) + H_{\text{leads}} + H_{\text{wire-leads}} \,.
\end{equation}
The wire is modeled by $N$ atomic orbitals $|n\rangle$, $n=1,\ldots,N$,
which are in a tight-binding description coupled by hopping matrix elements.
Then, the corresponding Hamiltonian for the electrons on the wire reads in
a second quantized form
\begin{equation}
H_{\text{wire}}(t)=\sum_{n,n'} H_{nn'}(t)\, c_n^\dagger c_{n'},
\end{equation}
where the fermionic operators $c_n$, $c_n^\dagger$ annihilate,
respectively create, an electron in the atomic orbital $|n\rangle$ and
obey the anti-commutation relation $[c_n,c_{n'}^\dagger]_+=\delta_{n,n'}$.
The influence of the laser field is given by a periodic time-dependence
of the on-site energies yielding a single particle Hamiltonian of the
structure $H_{nn'}(t)=H_{nn'}(t+\T)$, where
$\T=2\pi/\Omega$ is determined by the frequency $\Omega$ of the laser field.

The orbitals $|1\rangle$ and $|N\rangle$ at the left and the right end
of the molecule, that we shall term donor and acceptor, respectively,
are coupled to ideal leads (cf.\ Fig.~\ref{fig:wire}) by the tunneling
Hamiltonians
\begin{equation}
H_{\text{wire-leads}}
=\sum_q ( V_{qL} \, c_{qL}^\dagger c_1 + 
          V_{qR} \, c_{qR}^\dagger c_N) + \mathrm{H.c.}
\end{equation}
The operator $c_{qL}$ ($c_{qR}$) annihilates an electron in state $Lq$ ($Rq$)
on the left (right) lead.
The leads are modeled as non-interacting electrons with the
Hamiltonian
\begin{equation}
H_{\rm leads}=\sum_q(\epsilon_{qL}\, c_{qL}^\dagger c_{qL}+
\epsilon_{qR}\, c_{qR}^\dagger c_{qR}),
\end{equation}
where $\epsilon_{qL}$ is the single particle energy of the state $qL$
and correspondingly for the right lead.  As discussed above, the leads
can be described by a grand-canonical ensemble of electrons, i.e.\ by
a density matrix
\begin{equation}
\label{rholeadeq}
\varrho_\mathrm{leads,eq}\propto \exp\left[{-(H_\mathrm{leads}-\mu_L N_L
-\mu_R N_R)/k_BT}\right],
\end{equation}
where $\mu_{L/R}$ are the electro-chemical potentials and
$N_{L/R}=\sum_{q} c^\dagger_{qL/R} c_{qL/R}$ the electron numbers
in the left/right lead.  As a consequence, the only
non-trivial expectation values of lead operators read
\begin{equation}
\langle c_{qL}^\dagger c_{qL}\rangle = f(\epsilon_{qL}-\mu_{L}).
\end{equation}
Here, $f(x)=(1+e^{x/k_BT})^{-1}$ denotes the Fermi function.

\subsection{Perturbation theory}
While the leads and the wire, including the driving, will be treated
exactly, we take the wire-lead Hamiltonian as a perturbation into
account.  Starting from the Liouville-von Neumann equation $i \hbar
\dot \varrho(t)=[H(t),\varrho(t)]$  together with the factorizing initial condition
$\varrho(t_0) = \varrho_\mathrm{wire}(t_0)\otimes
\varrho_\mathrm{leads,eq}$, we derive by standard techniques an
approximate equation of motion for the total density operator
$\varrho(t)$.  This is most conveniently achieved in
the interaction picture with respect to the uncoupled dynamics where
the Liouville-von Neumann equation reads
\begin{equation}
i\hbar\frac{d}{dt}\tilde\varrho(t,t_0)
=[\widetilde H_{\mathrm{wire-leads}}(t,t_0),\tilde\varrho(t,t_0)] .
\label{LvN}
\end{equation}
The tilde denotes the corresponding interaction picture operators,
$\widetilde X(t,t')=U_0^\dagger(t,t')\,X(t)\,U_0(t,t')$ where the propagator of
the wire and the lead in the absence of the lead-wire coupling is given by the
time-ordered product
\begin{equation}
U_0(t,t') = {\stackrel{\leftarrow}{\textstyle T}}
\exp\left(-\frac{i}{\hbar}\int_{t'}^t
dt''\,[H_\mathrm{wire}(t'')+H_\mathrm{leads}]\right) .
\end{equation}
Equation \eqref{LvN} is equivalent to the following integral equation
\begin{equation}
\tilde\varrho(t,t_0)=\tilde\varrho(t_0,t_0)-\frac{i}{\hbar}\int_{t_0}^t dt'
[\widetilde H_{\mathrm{wire-leads}}(t',t_0),\tilde\varrho(t',t_0)] .
\label{LvN,int}
\end{equation}
We reinsert this expression into the differential equation \eqref{LvN}
and use that to zeroth order in the molecule-lead coupling the
interaction-picture density operator does not change with time,
$\tilde\varrho(t-\tau,t_0)\approx\tilde\varrho(t,t_0)$.  A
transformation back to the Schr\"odinger picture results in the
following approximate equation of motion for the total density
operator \cite{Lehmann2002b,Lehmann2002d}
\begin{equation}
\begin{split}
\label{mastereq}
\dot\varrho(t) =
& -\frac{i}{\hbar}[H_{\rm wire}(t)+H_{\rm leads},\varrho(t)]
  -\frac{1}{\hbar^2}\int\limits_0^{\infty} \!\!d\tau
  [H_\mathrm{wire-leads},[\widetilde H_\mathrm{wire-leads}(t-\tau,t),
  \varrho(t)]] \\
& -\frac{i}{\hbar}[H_\mathrm{wire-leads},
  U_0(t,t_0)\varrho(t_0)U_0^\dagger(t,t_0)]
.
\end{split}
\end{equation}
Since we only consider asymptotic times $t-t_0\to\infty$, we have set
the upper limit in the integral to infinity.  The third term in
Eq.~\eqref{mastereq} stems from the initial condition at $t_0$ in the
integrated form \eqref{LvN,int} of the Liouville-von Neumann equation.
For the chosen factorizing initial condition, it will not contribute
to the expectation values calculated below.

The net (incoming minus outgoing) current through the left contact is given by
the negative time derivative of the electron number in the left lead,
multiplied by the electron charge $-e$, i.e.\
\begin{equation}
I_L(t)  =e \mathop{\rm tr}[\dot \varrho(t) N_L] .
\end{equation}
We insert $\dot\varrho(t)$ from Eq.~\eqref{mastereq} and obtain an
expression that depends on the density of states in the leads times
their coupling strength to the connected sites.
At this stage it is convenient to introduce the spectral density of the
lead-wire coupling
\begin{equation}
\label{WBL}
\Gamma_{L/R}(\epsilon) = \frac{2\pi}{\hbar}
\sum_q |V_{qL/R}|^2 \delta(\epsilon-\epsilon_{qL/R}),
\end{equation}
which fully describes the leads' influence.  If the lead states are
dense, $\Gamma_{L/R}(\epsilon)$ becomes a continuous function.
Because we
are mainly interested in the behavior of the molecule and not in the details
of the lead-wire coupling, we assume that the conduction band width of the
leads is much larger than all remaining relevant energy scales.  Consequently,
we approximate in the so-called wide-band limit the functions
$\Gamma_{L/R}(\epsilon)$ by the constant values $\Gamma_{L/R}$.
After some algebra, we find for the \textit{time-dependent} net electrical
current through the left contact the expression
\begin{equation}
I_L(t)  = 
\frac{e\Gamma_L}{\pi\hbar}\mathop{\rm Re}\int\limits_0^\infty \!d\tau \! \int
\!d\epsilon\, e^{i\epsilon\tau/\hbar}
\Big\{
\big\langle \tilde c_1^\dagger(t-\tau,t)\, c_1\big\rangle
-[c_1,\tilde c_1^\dagger(t-\tau,t)]_+ f(\epsilon-\mu_L)
\Big\} ,
\label{current_general}
\end{equation}
and correspondingly for the current through the contact on the
right-hand side. Here, we made the assumption, that the leads are at
all times well described by the density operator~\eqref{rholeadeq}.
Note that the anti-commutator $[c_1,\tilde c_1^\dagger(t-\tau,t)]_+$
is in fact a c-number.  Like the expectation value $\langle
c_1^\dagger(t-\tau,t)\, c_1\rangle$ it depends on the dynamics of the
isolated wire and is influenced by the external driving.
The first contribution of the $\epsilon$-integral in
Eq.~\eqref{current_general} is readily evaluated to yield an expression
proportional to $\delta(\tau)$.  Thus, this term becomes local in time and
reads $e\Gamma_L\big\langle c_1^\dagger c_1\big\rangle$.

\subsection{Floquet decomposition}
Let us next focus on the single-particle dynamics of the driven molecule
decoupled from the leads.
Since its Hamiltonian is periodic in time, $H_{nn'}(t)=H_{nn'}(t+\T)$,
we can solve the corresponding time-dependent Schr\"odinger
equation within a Floquet approach.  This means that we make use of the fact
that there exists a complete set of solutions of the form
\cite{Shirley1965a,Sambe1973a,Fainshtein1978a,Hanggi1998a, Grifoni1998a}
\begin{equation}
|\Psi_\alpha(t)\rangle=e^{-i\epsilon_\alpha t/\hbar}
|\Phi_\alpha(t)\rangle,\ |\Phi_\alpha(t)\rangle=|\Phi_\alpha(t+\mathcal{T})\rangle
\end{equation}
with the quasi-energies $\epsilon_\alpha$.
Since the so-called Floquet modes $|\Phi_\alpha(t)\rangle$ obey the
time-periodicity of the driving field, they can be decomposed into the
Fourier series
\begin{equation}
\label{floquetseries}
|\Phi_\alpha(t)\rangle=\sum_k e^{-ik\Omega t}
|\Phi_{\alpha,k}\rangle.
\end{equation}
This implies that the quasienergies $\epsilon_\alpha$ come in classes,
\begin{equation}
\epsilon_{\alpha,k}=\epsilon_\alpha+k\hbar\Omega,\quad k=0,\pm1, \pm2,\ldots,
\end{equation}
of which all members represent the same solution of the Schr\"odinger equation.
Therefore, the quasienergy spectrum can be reduced to a single
``Brillouin zone'' $-\hbar\Omega/2\leq \epsilon<\hbar\Omega/2$.
In turn, all physical quantities that are computed within a Floquet
formalism are independent of the choice of a specific class member.
Thus, a consistent description must obey the so-called class invariance, i.e.\
it must be invariant under the substitution
of one or several Floquet states by equivalent ones,
\begin{equation}
\label{classinvariance}
\epsilon_\alpha,\,|\Phi_\alpha(t)\rangle \longrightarrow
\epsilon_\alpha+k_\alpha\hbar\Omega,\,
e^{ik_\alpha\Omega t}|\Phi_\alpha(t)\rangle ,
\end{equation}
where $k_1, \dots, k_N$ are integers.
In the Fourier decomposition \eqref{floquetseries}, the prefactor
$\exp({ik_\alpha\Omega t})$ corresponds to a shift of the side band
index so that the class invariance can be expressed equivalently as
\begin{equation}
\label{classinvariance_k}
\epsilon_\alpha,\,|\Phi_{\alpha,k}\rangle \longrightarrow
\epsilon_\alpha+k_\alpha\hbar\Omega,\,|\Phi_{\alpha,k+k_\alpha}\rangle .
\end{equation}
Floquet states and quasienergies can be obtained from the quasienergy equation
\cite{Shirley1965a, Sambe1973a, Fainshtein1978a, Manakov1986a, Hanggi1998a,
Grifoni1998a}
\begin{equation}
\label{floquet_hamiltonian}
\Big(\sum_{n,n'}|n\rangle H_{nn'}(t) \langle n'|-i\hbar\frac{d}{dt}\Big)
|\Phi_{\alpha}(t)\rangle = \epsilon_\alpha |\Phi_{\alpha}(t)\rangle .
\end{equation}
A wealth of methods for the solution of this eigenvalue problem can be found
in the literature.  For an overview, we refer the reader to the
reviews in Refs.~~\onlinecite{Hanggi1998a,Grifoni1998a}, and the references therein.

As the equivalent of the one-particle Floquet states $|\Phi_\alpha(t)\rangle$,
we define a Floquet picture for the fermionic creation and annihilation
operators $c_n^\dagger$, $c_n$, by the time-dependent transformation
\begin{equation}
c_\alpha(t) = \sum_n \langle\Phi_\alpha(t)|n\rangle\, c_n . \label{c_alpha}
\end{equation}
The inverse transformation
\begin{equation}
c_n = \sum_{\alpha} \langle n|\Phi_\alpha(t)\rangle\,c_\alpha(t)
\label{c_n}
\end{equation}
follows from the mutual orthogonality and the completeness of the
Floquet states at equal times \cite{Hanggi1998a, Grifoni1998a}.  Note
that the right-hand side of Eq.~\eqref{c_n} becomes $t$-independent
after the summation.  The operators $c_\alpha(t)$ are constructed in such a
way that the time-dependences of the interaction picture operators
$\tilde c_\alpha(t-\tau,t)$ separate, which will turn out to be crucial for 
the further analysis. Indeed, one can easily verify the relation
\begin{equation}
  \begin{split}
    \tilde c_\alpha(t-\tau,t)
    &=U_0^\dagger(t-\tau,t)\,c_\alpha(t-\tau)\,U_0(t-\tau,t) \\
    &=e^{i\epsilon_\alpha \tau/\hbar} c_\alpha(t)  
    \label{c_alpha_tilde}    
  \end{split}
\end{equation}
by differentiating the definition in the first line
with respect to $\tau$ and using that
$|\Phi_\alpha(t)\rangle$ is a solution of the eigenvalue equation
\eqref{floquet_hamiltonian}. The fact that the initial condition $\tilde
c_\alpha(t,t)=c_\alpha(t)$ is fulfilled completes the proof.
The corresponding expression for the interaction picture operator in
the on-site basis, $\tilde c_n(t-\tau,t)$, can be derived with help
of Eq.~\eqref{c_n} at time $t-\tau$ together with \eqref{c_alpha_tilde}
to read
\begin{align}
\tilde c_n(t-\tau,t)
&= \sum_\alpha\langle n|\Phi_\alpha(t-\tau)\rangle
   e^{i\epsilon_\alpha\tau/\hbar} c_\alpha(t) \\
&= \sum_{\alpha k} e^{i(\epsilon_\alpha/\hbar+k\Omega)\tau}
    e^{-ik\Omega t}\langle n|\Phi_{\alpha,k}\rangle c_\alpha(t) .
\label{c_n_tilde}
\end{align}
Equations~\eqref{c_alpha_tilde}, \eqref{c_n_tilde}, consequently allow to
express the interaction picture operator $\tilde c_1^\dagger(t-\tau,t)$
appearing in the current formula \eqref{current_general} via $c_\alpha(t)$,
dressed by exponential prefactors.  

This spectral decomposition allows one to carry out the time and
energy integrals in the expression \eqref{current_general} for the net
current entering the wire from the left lead.  Thus, we obtain
\begin{equation}
\label{I_L(t)}
I_L(t) = \sum_k e^{-ik\Omega t}I_L^k ,
\end{equation}
with the corresponding Fourier components
\begin{equation}
\begin{split}
I_L^k = {} &
e\Gamma_L
\bigg[
\sum_{\alpha\beta k' k''}
 \langle \Phi_{\alpha, k'+k''} |1\rangle\langle 1|\Phi_{\beta, k+k''} \rangle
 R_{\alpha\beta,k'}\\
& \qquad-\frac{1}{2}\sum_{\alpha k'}
\Big(
  \langle \Phi_{\alpha, k'}|1\rangle \langle 1|\Phi_{\alpha, k+k'}\rangle 
    +\langle \Phi_{\alpha, k'-k}|1\rangle \langle 1|\Phi_{\alpha,
    k'}\rangle 
\Big) f(\epsilon_{\alpha,k'}-\mu_L) 
\bigg] .
\end{split}
\label{I_fourier}
\end{equation}
Here, we have introduced the expectation values
\begin{align}
R_{\alpha\beta}(t)&=\langle c_\alpha^\dagger(t) c_\beta(t)\rangle
 =R_{\beta\alpha}^*(t) \\
&=\sum_k e^{-ik\Omega t}R_{\alpha\beta,k}.
\end{align}
The Fourier decomposition in the last line is possible because all
$R_{\alpha\beta}(t)$ are expectation values of a linear, dissipative,
periodically driven system and therefore share in the long-time limit
the time-periodicity of the driving field.
In the subspace of a single electron, $R_{\alpha\beta}$ reduces to
the density matrix in the basis of the Floquet states which has
been used to describe dissipative driven quantum systems
in Refs.~~\onlinecite{Blumel1991a, Dittrich1993a, Kohler1997a, Kohler1998a,
Grifoni1998a, Hanggi2000a}.

The next step towards the stationary current is to find the Fourier
coefficients $R_{\alpha\beta,k}$ at asymptotic times.  To this end,
we derive from the equation of motion \eqref{mastereq} a
master equation for $R_{\alpha\beta}(t)$.  Since all coefficients
of this master equation, as well as its asymptotic solution, are
$\T$-periodic, we can split it into its Fourier components.
Finally, we obtain for the $R_{\alpha\beta,k}$ the inhomogeneous set of
equations
\begin{align}
\label{mastereq_fourier}
{\frac{i}{\hbar}(\epsilon_\alpha-\epsilon_\beta+k \hbar \Omega)R_{\alpha\beta,k}}
=&
 \frac{\Gamma_L}{2}\sum_{k'}
    \Big( 
    \sum_{\beta'k''}
                \langle\Phi_{\beta,k'+k''}|1\rangle
                \langle 1|\Phi_{\beta',k+k''}\rangle
                R_{\alpha\beta',k'}
\\& \hspace{6ex} +
    \sum_{\alpha'k''}
                \langle\Phi_{\alpha',k'+k''}|1\rangle
                \langle 1|\Phi_{\alpha,k+k''}\rangle
                R_{\alpha'\beta,k'}
\nonumber \\ & \hspace{6ex} - 
    \langle\Phi_{\beta,k'-k}|1\rangle
    \langle 1|\Phi_{\alpha,k'}\rangle
    f(\epsilon_{\alpha,k'}-\mu_L)
\nonumber \\ & \hspace{6ex} - 
    \langle\Phi_{\beta,k'}|1\rangle
    \langle 1|\Phi_{\alpha,k'+k}\rangle
    f(\epsilon_{\beta,k'}-\mu_L)
   \Big)
\nonumber \\ &\hspace{-0ex} {} +
  \text{same terms with the replacement}
\nonumber \\ &\hspace{2ex} {}
  {\big\{\Gamma_L, \mu_L, |1\rangle\langle 1|\big\} \rightarrow
   \big\{\Gamma_R, \mu_R, |N\rangle\langle N|\big\}}.
\nonumber
\end{align}
For a consistent Floquet description, the current formula together with
the master equation must obey class invariance.
Indeed, the simultaneous transformation with \eqref{classinvariance_k}
of both the master equation \eqref{mastereq_fourier} and the current
formula \eqref{I_fourier} amounts to a mere shift of summation indices and,
thus, leaves the current as a physical quantity unchanged.

For the typical parameter values used below, a large number of sidebands
contributes significantly to the Fourier decomposition of the Floquet modes
$|\Phi_{\alpha}(t)\rangle$.  Numerical convergence for the solution of the
master equation \eqref{mastereq_fourier}, however, is already obtained by just
using a few sidebands for the decomposition of $R_{\alpha\beta}(t)$.  This
keeps the numerical effort relatively small and justifies \textit{a posteriori}
the use of the Floquet representation \eqref{c_n}. Yet we are able to treat
the problem beyond the rotating-wave-approximation.

\subsection{Time-averaged current through the molecular wire}

Equation \eqref{I_L(t)} implies that the current $I_L(t)$ obeys the
time-periodicity of the driving field. Since we consider here excitations by a
laser field, the corresponding driving frequency lies in the optical or
infrared spectral range. In an experiment one will thus only be able to
measure the time-average of the current.  For the net current entering through
the left contact it is given by
\begin{equation}
\label{dc_current_l}
\begin{split}
\bar{I}_L =  I^0_L = 
e\Gamma_L\sum_{\alpha k}\Big[&
\sum_{\beta k'}
 \langle \Phi_{\alpha, k'+k} |1\rangle\langle 1|\Phi_{\beta, k'} \rangle
 R_{\alpha\beta,k}
-
\langle \Phi_{\alpha, k}|1\rangle \langle 1|\Phi_{\alpha, k}\rangle
  f(\epsilon_{\alpha,k}-\mu_L)
\Big] .
\end{split}
\end{equation}
By replacing $\{|1\rangle,L\}\to\{|N\rangle,R\}$, one obtains for the current
which enters from the right, $I_R(t)$, and the corresponding Fourier
coefficients and time averages.

Total charge conservation of the original wire-lead Hamiltonian
\eqref{wire-lead-hamiltonian} of course requires that the charge on
the wire can only change by current flow, amounting to the continuity
equation $\dot Q_{\mathrm{wire}}(t)=I_L(t)+I_R(t)$.  Since
asymptotically, the charge on the wire obeys at most the periodic
time-dependence of the driving field, the time-average of $\dot
Q_{\mathrm{wire}}(t)$ must vanish in the long-time limit.  From the
continuity equation one then finds that $\bar I_L + \bar
I_R=0$, and we can introduce the time-averaged current
\begin{equation}
\bar I = \bar I_L = -\bar I_R.
\label{barI}
\end{equation}
This continuity equation can be obtained directly from the average current
formula \eqref{dc_current_l} together with the master equation
\eqref{mastereq_fourier}, as has been explicitly shown in
Ref.~~\onlinecite{Lehmann2002d}.

\section{Laser-enhanced current}
\label{sec:application}

\subsection{Bridged molecular wire}

As a working model we consider a molecule consisting of a donor and an acceptor
site and $N-2$ sites in between (cf.\ Fig.~\ref{fig:wire}).  Each of the $N$
sites is coupled to its nearest neighbors by a hopping matrix elements
$\Delta$.  The laser field renders each level oscillating in time with a
position-dependent amplitude.  Thus, the corresponding time-dependent wire
Hamiltonian reads
\begin{equation}
\begin{split}
H_{nn'}(t)= &
-\Delta(\delta_{n,n'+1}+\delta_{n+1,n'})
+ \left[ E_n - A\,x_n\cos(\Omega t)\right]\delta_{nn'},
\label{wirehamiltonian}
\end{split}
\end{equation}
where $x_n=(N+1-2n)/2$ is the scaled position of site $|n\rangle$.  The energy
$a(t)$ equals the electron charge multiplied by the electrical field amplitude
of the laser and the distance between two neighboring sites. The energies
of the donor and the acceptor orbitals are assumed to be at the level of the
chemical potentials of the attached leads, $E_1=E_N=\mu_L=\mu_R$.  The bridge
levels $E_n$, $n=2,\dots,N-1$, lie $E_B\gg\Delta$ above the chemical
potential, as sketched in Figure~\ref{fig:wire}.

In all numerical studies, we will use a symmetric coupling,
$\Gamma_L=\Gamma_R=\Gamma$.  The hopping matrix element $\Delta$ serves as the
energy unit; in a realistic wire molecule, $\Delta$ is of the order $0.1\,{\rm
eV}$.  Thus, our chosen wire-lead hopping rate $\Gamma=0.1 \Delta/\hbar$
yields $e\Gamma=2.56\times10^{-5}$\,Amp\`ere and $\Omega\approx10\Delta/\hbar$
corresponds to a laser frequency in the near infrared.  For a typical distance
of {$5$\AA} between two neighboring sites, a driving amplitude $A=\Delta$ is
equivalent to an electrical field strength of $2\times10^6\,\mathrm{V/cm}$.

\subsection{Average current at resonant excitations}

\begin{figure}
\includegraphics[width=.5\columnwidth]{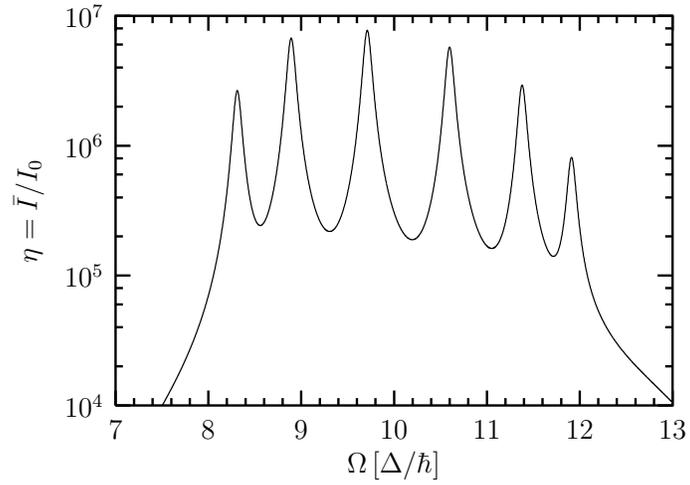}
\caption{ \label{fig:peaks}
Amplification of the 
time-averaged current through the wire sketched in Figure~\ref{fig:wire} with
$E_B=10\Delta$.  The scaled amplitude is $A=0.1\Delta$; the applied voltage
$\mu_L-\mu_R=5\Delta/e$.
The other parameters read $\Gamma=\Gamma_L=\Gamma_R=0.1\Delta/\hbar$,
$k_BT=0.25\Delta$.  }
\end{figure}%
\begin{figure}
\includegraphics[width=.5\columnwidth]{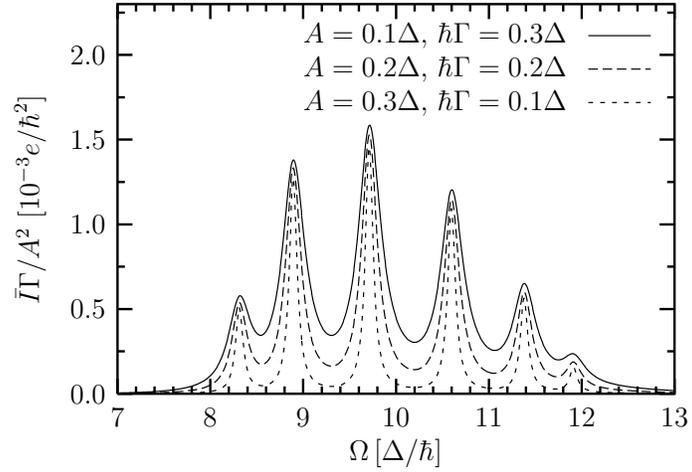}
\caption{ \label{fig:scaling_f}
Average current $\bar I$ as a function of the the driving frequency
$\Omega$ for various driving amplitudes $A$ and coupling
strength~$\Gamma=\Gamma_L=\Gamma_R$.
All the other parameters are as in Fig.~\ref{fig:peaks}.}
\end{figure}%
\begin{figure}
\includegraphics[width=.5\columnwidth]{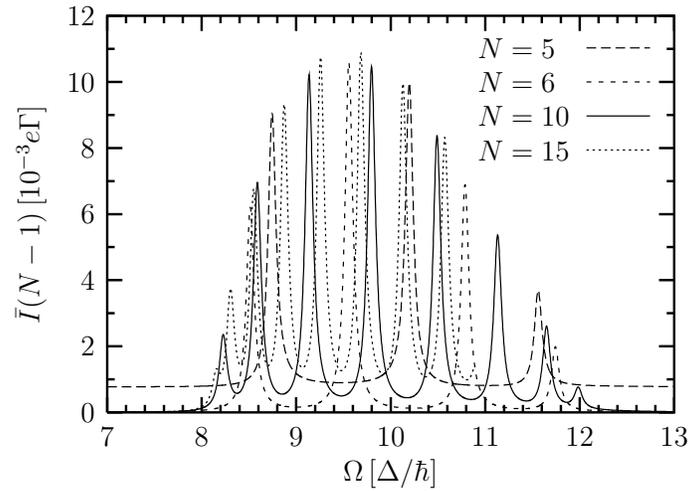}
\caption{ \label{fig:scaling_N}
Average current $\bar I$ as a function of the the driving frequency
$\Omega$ for various wire length $N$.  All the other parameters are as in
Fig.~\ref{fig:peaks}.}
\end{figure}%
%
Let us first discuss the static problem in the absence of the field, i.e.\ for
$A=0$.  In the present case where the coupling between two neighboring sites
is much weaker than the bridge energy, $\Delta\ll E_B$, one finds two types of
eigenstates: One forms a doublet whose states are approximately given by
$(|1\rangle\pm|N\rangle)/\sqrt{2}$.  Its splitting can be estimated in a
perturbational approach \cite{Ratner1990a} and is approximately given by
$2\Delta(\Delta/E_B)^{N-2}$.  A second group of states is located on the
bridge.  It consists of $N-2$ levels with energies in
the range $[E_B-2\Delta,E_B+2\Delta]$.  In the absence of the driving field,
these bridge states mediate the super-exchange between the donor and the
acceptor.  This yields an exponentially decaying length dependence of the
conductance \cite{Mujica1994a,Nitzan2001a}.

This behavior changes significantly when a driving field with a frequency
$\Omega\approx E_B/\hbar$ is switched on.  Then the resonant bridge levels
merge with the donor and the acceptor state to form a Floquet state.  This
opens a direct channel for the transport resulting in an enhancement of the
electron current as depicted in Figure~\ref{fig:peaks} where we plot the
current amplification, defined as the ratio of the time-averaged current to
the current in the absence of the laser, $\eta=\bar I/I_0$: In a wire with
$N=8$ sites, one finds peaks in the current when the driving frequency matches
the energy difference between the donor/acceptor doublet and one of the
$N-2=6$ bridge levels.  The applied voltage is always chosen so small that the
bridge levels lie below the chemical potentials of the leads.  The
amplification, can assume many orders of magnitude, cf.\
Figure~\ref{fig:peaks}.  Generally, the response of a system to a weak
resonant driving scales with the damping and the driving amplitude.  Figure
\ref{fig:scaling_f} demonstrates this behavior for the peaks of the electrical
current.  The peak heights at the maxima of the time-averaged current are
found proportional to $A^2/\Gamma$.  A further scaling behavior is found for
the current peaks as a function of the wire length:  The average current no
longer exhibits the exponentially decaying length dependence that has been
found for bridged super-exchange.  By contrast, it emerges proportional to
$1/(N-1)$.  This can be appreciated in Figure~\ref{fig:scaling_N} where the
scale of the abscissa is chosen proportional to $N-1$ such that it suggests a
common envelope function.  Put differently, the current is essentially
inversely proportional to the length as in the case of Ohmic conductance.

In summary, we find current peaks whose height $\bar I_\mathrm{peak}$ scales
according to
\begin{equation}
\label{scaling}
\bar I_\mathrm{peak} \propto \frac{A^2}{(N-1)\Gamma}.
\end{equation}
Thus, the current is especially for long wires much larger than the
corresponding current in the absence of the driving.

\section{Conclusions}

We have presented a detailed derivation of the Floquet transport formalism
which has been applied in Refs.~~\onlinecite{Lehmann2002b, Lehmann2002c,
Lehmann2002d}.  The analysis of a bridged molecular wire revealed that
resonant excitations from the levels that connect the molecule to the external
leads to bridge levels yield peaks in the current as a function of the driving
frequency.  In a regime with weak driving and weak electron-lead coupling,
$\Delta\gg\Gamma,A$, the peak heights scale with the coupling strength, the
driving amplitude, and the wire length.  The laser irradiation induces a large
current enhancement of several orders of magnitude.  The observation of these
resonances could serve as an experimental starting point for the more
challenging attempt of measuring quantum ratchet effects \cite{Lehmann2002b,
Lehmann2002d} or current switching by laser fields \cite{Lehmann2002c}.

\section{acknowledgement}

This work has been supported by SFB 486
and by the Volkswagen-Stiftung under grant No.~I/77~217.
One of us (S.C.) has been supported by a European Community Marie Curie
Fellowship.

\bibliographystyle{prsty_ph}

\begin{thebibliography}{10}

\bibitem{Aviram1974a}
A. Aviram and M.~A. Ratner, Chem. Phys. Lett. {\bf 29},  277  (1974).

\bibitem{Joachim2000a}
C. Joachim, J.~K. Gimzewski, and A. Aviram, Nature {\bf 408},  541  (2000).

\bibitem{Nitzan2001a}
A. Nitzan, Annu. Rev. Phys. Chem. {\bf 52},  681  (2001).

\bibitem{Hanggi2002a}
{P.~H\"anggi, M.~Ratner, and S.~Yaliraki, Special Issue: Processes in Molecular
  Wires, Chem. Phys. \textbf{281}, pp. 111-502 (2002).}

\bibitem{Cui2001a}
X.~D. Cui {\it et~al.}, Science {\bf 294},  571  (2001).

\bibitem{Reichert2002a}
J. Reichert, R. Ochs, D. Beckmann, H.~B. Weber, M. Mayor, and H.
  v.~L\"ohneysen, Phys. Rev. Lett. {\bf 88},  176804  (2002).

\bibitem{Lehmann2002b}
J. Lehmann, S. Kohler, P. H\"anggi, and A. Nitzan, Phys. Rev. Lett. {\bf 88},
  228305  (2002).

\bibitem{Lehmann2002d}
J. Lehmann, S. Kohler, P. H\"anggi, and A. Nitzan, cond-mat/  0208404  (2002).

\bibitem{Lehmann2002c}
J. Lehmann, S. Camalet, S. Kohler, and P. H\"anggi, physics/  0205060  (2002).

\bibitem{Mujica1994a}
V. Mujica, M. Kemp, and M.~A. Ratner, J. Chem. Phys. {\bf 101},  6849  (1994).

\bibitem{Datta1995a}
S. Datta, {\em Electronic Transport in Mesoscopic Systems} (Cambridge
  University Press, Cambridge, 1995).

\bibitem{Nitzan2001b}
A. Nitzan, J. Phys. Chem. A {\bf 105},  2677  (2001).

\bibitem{Petrov2001a}
E.~G. Petrov and P. H\"anggi, Phys. Rev. Lett. {\bf 86},  2862  (2001).

\bibitem{Lehmann2002a}
J. Lehmann, G.-L. Ingold, and P. H\"anggi, Chem. Phys. {\bf 281},  199  (2002).

\bibitem{Manakov1986a}
N.~L. Manakov, V.~D. Ovsiannikov, and L.~P. Rapoport, Phys. Rep. {\bf 141},
  319  (1986).

\bibitem{Grifoni1998a}
M. Grifoni and P. H\"anggi, Phys. Rep. {\bf 304},  229  (1998).

\bibitem{Blumel1989a}
R. Bl\"umel, R. Graham, L. Sirko, U. Smilansky, H. Walther, and K. Yamada,
  Phys. Rev. Lett. {\bf 62},  341  (1989).

\bibitem{Kohler1997a}
S. Kohler, T. Dittrich, and P. H\"anggi, Phys. Rev. E {\bf 55},  300  (1997).

\bibitem{Shirley1965a}
J.~H. Shirley, Phys. Rev. {\bf 138},  B979  (1965).

\bibitem{Sambe1973a}
H. Sambe, Phys. Rev. A {\bf 7},  2203  (1973).

\bibitem{Fainshtein1978a}
A.~G. Fainshtein, N.~L. Manakov, and L.~P. Rapoport, J. Phys. B {\bf 11},  2561
   (1978).

\bibitem{Hanggi1998a}
P. H\"anggi,  in {\em Quantum Transport and Dissipation} (Wiley-VCH, Weinheim,
  1998).

\bibitem{Blumel1991a}
R. Bl\"umel, A. Buchleitner, R. Graham, L. Sirko, U. Smilansky, and H. Walter,
  Phys. Rev. A {\bf 44},  4521  (1991).

\bibitem{Dittrich1993a}
T. Dittrich, B. Oelschl\"agel, and P. H\"anggi, Europhys. Lett. {\bf 22},  5
  (1993).

\bibitem{Kohler1998a}
S. Kohler, R. Utermann, P. H\"anggi, and T. Dittrich, Phys. Rev. E {\bf 58},
  7219  (1998).

\bibitem{Hanggi2000a}
P. H\"anggi, S. Kohler, and T. Dittrich,  in {\em Statistical and Dynamical
  Aspects of Mesoscopic Systems}, Vol.~547 of {\em Lecture Notes in Physics},
  edited by D. Reguera, G. Platero, L.~L. Bonilla, and J.~M. Rub\'{\i}
  (Springer, Berlin, 2000), pp.\ 125--157.

\bibitem{Ratner1990a}
M.~A. Ratner, J. Phys. Chem. {\bf 94},  4877  (1990).

\end{thebibliography}


\end{document}